\documentclass[modern]{aastex62}

\graphicspath{{./}{figures/}}

\accepted{November 09, 2020}
\submitjournal{ApJ}

\shorttitle{Imaging Spectroscopy of CME-Associated Solar Radio Bursts}
\shortauthors{Chhabra et al.}

\begin{document}

\title{Imaging Spectroscopy of CME-Associated Solar Radio Bursts using OVRO-LWA}

\correspondingauthor{Sherry Chhabra}
\email{sc787@njit.edu}

\author[0000-0001-7754-0804]{Sherry Chhabra}
\affil{Center for Solar-Terrestrial Research, New Jersey Institute of Technology, Newark, NJ 07102, USA}
\affil{NASA Goddard Space Flight Center, Greenbelt, MD 20771, USA}
\author[0000-0003-2520-8396]{Dale E. Gary}
\affiliation{Center for Solar-Terrestrial Research, New Jersey Institute of Technology, Newark, NJ 07102, USA}

\author[0000-0002-7083-4049]{Gregg Hallinan}
\affiliation{California Institute of Technology, 1200 E California Blvd MC 249-17, Pasadena, CA 91125, USA}
\affiliation{Owens Valley Radio Observatory, MC 249-17 California Institute of Technology, Pasadena, CA 91125, USA}

\author[0000-0003-2238-2698]{Marin M. Anderson}
\affiliation{California Institute of Technology, 1200 E California Blvd MC 249-17, Pasadena, CA 91125, USA}
\affiliation{Owens Valley Radio Observatory, MC 249-17 California Institute of Technology, Pasadena, CA 91125, USA}

\author[0000-0002-0660-3350]{Bin Chen}
\affiliation{Center for Solar-Terrestrial Research, New Jersey Institute of Technology, Newark, NJ 07102, USA}

\newcommand{\SWIN}{Centre for Astrophysics \& Supercomputing, Swinburne University of Technology, Hawthorn, VIC 3122, Australia}

\newcommand{\CFA}{Harvard-Smithsonian Center for Astrophysics, 60 Garden Street, Cambridge, MA 02138, USA}

\author[0000-0003-4912-5974]{Lincoln J.\ Greenhill}
\affiliation{\CFA}

\author[0000-0003-2783-1608]{Danny C.\ Price}
\affiliation{\CFA}
\affiliation{\SWIN}

\begin{abstract}
 We present first results of a solar radio event observed with the Owens Valley Radio Observatory Long Wavelength Array (OVRO-LWA) at metric wavelengths. We examine a complex event consisting of multiple radio sources/bursts associated with a fast coronal mass ejection (CME) and an M2.1 GOES soft X-ray flare from 2015 September 20. Images of 9--s cadence are used to analyze the event over a 120-minute period, and solar emission is observed out to a distance of $\approx3.5\,R_\odot$, with an instantaneous bandwidth covering 22~MHz within the frequency range of 40--70~MHz. We present our results from the investigation of the radio event, focusing particularly on one burst source that exhibits outward motion, which we classify as a moving type IV burst. We image the event at multiple frequencies and use the source centroids to obtain the velocity for the outward motion. Spatial and temporal comparison with observations of the CME in white light from the LASCO(C2) coronagraph, indicates an association of the outward motion with the core of the CME.  By performing graduated-cylindrical-shell (GCS) reconstruction of the CME, we constrain the density in the volume. The electron plasma frequency obtained from the density estimates do not allow us to completely dismiss plasma emission as the underlying mechanism. However, based on source height and smoothness of the emission in frequency and time, we argue that gyrosynchrotron is the more plausible mechanism. We use gyrosynchrotron spectral fitting techniques to estimate the evolving physical conditions during the outward motion of this burst source. 
\end{abstract}

)

\section{Introduction} \label{sec:Intro}

Solar radio bursts in the metric wavelength range can be produced via interactions of plasma waves and particles (generically called plasma radiation), or directly via energetic nonthermal particles that gyrate in the ambient magnetic field (gyroradiation).  Since their discovery in the 1950s \citep{1950AuSRA...3..387W}, solar radio bursts have been the subject of intense investigation. Radio diagnostics
hold great importance for characterizing the space environment at the Sun, near Earth, and elsewhere in the solar system. Therefore the study of these phenomena, both spatially and spectrally, contributes significantly to our understanding of the properties of the ambient medium, the instabilities that may cause them, and subsequently, the underlying fundamental processes of electron acceleration and magnetic reconnection involved in impulsive phenomena such as flares and coronal mass ejections (CMEs).

At metric wavelengths, solar type II, type III, and type IV radio bursts are often associated with impulsive events \citep{Sakurai1974, White07, gopalswamy2011}.
Type III bursts are transient bursts known for their high rates of frequency drift, believed to be associated with electron beams traveling through plasma with a density gradient. Propagating electron beams of semi-relativistic ($\approx\,$0.1--0.5\,$c$) speeds can produce a bump-on-tail instability that generates Langmuir waves at the local plasma frequency, \(\textup{$\nu_p$} = 8980\,\sqrt[]{n(r)}\) Hz, where $n(r)$ is the electron density of the medium (in cm$^{-3}$) as a function of radial distance $r$ from the solar surface \citep{Ginzburg58}. The non-linear particle-wave and wave-wave interaction produces electromagnetic emission at the plasma frequency and its second harmonic respectively. Group type IIIs and complex type-III-like fast-drift bursts that occur with coronal mass ejections (CMEs) have been claimed to originate variously from shock-accelerated electrons, unspecified ``shock-associated'' acceleration, or acceleration directly from the flare site \citep [see][for details]{Cane81, Kundu84, Dulk2000, 2000ApJ...530...1049, Gopalswamy2010}. This uncertainty may be resolved with recently available high spatial, spectral, and temporal resolution imaging of type III bursts, which provide key information about reconnection sites and contribute to our understanding of particle acceleration \citep{Chen2013, Chen2018, Mulay2019}.

Metric-decametric bursts of type IV present themselves as broadband continuum emission in the dynamic spectrum, sometimes accompanied by fine structures like Zebra patterns, fiber bursts or broadband quasi-periodic pulsations \citep{Slottje1981}. They are further classified into stationary type IV bursts and moving type IVm bursts \citep{weiss63}.  Type IVm bursts exhibit a relatively slow increase in height with time. There is a general consensus that these bursts are generated by electrons trapped inside CMEs \citep{boischot68,smerd71}. Observations of moving type IV bursts associated with CMEs are rare--only 5\% of CMEs exhibit type IVm bursts according to \cite{1986SoPh..104..175G}. To our knowledge only 4 studies have so far reported moving type IV bursts in solar cycle 24 \citep{Tun13, bain2014, morosan2019,  vasanth2019}. Although immense progress has been made in understanding these bursts over the decades, the underlying mechanisms responsible for their production remain uncertain. \cite{Bastian2001} and \cite{Maia07} could clinch the case for gyrosynchrotron emission in their events because smooth continuum emission could be seen throughout the associated CME loops, although it should be noted that some areas of gyrosynchrotron emission in the interior of the CMEs were far brighter. Only this brighter interior emission, coincident with the core of the CME, could be seen in the moving type IV events reported by \cite{Tun13} \& \cite{bain2014}, but both argued for a gyrosynchrotron interpretation due to the smoothness of the emission in frequency and time. However, a recent study by \cite{morosan2019} found that a coherent emission mechanism (either plasma or electron cyclotron maser emission) was responsible for the moving type IV burst in their event, while a stationary component during an earlier time in the burst was found to be consistent with gyrosynchrotron emission \citep[see also][]{carley2017}. Therefore, it seems clear that some type IV and IVm continuum may be due to either gyrosynchrotron emission or a coherent mechanism or some combination \citep{gary1985, morosan2019, vasanth2019}. Given these alternatives, some effort is required to investigate the relative likelihood of gyrosynchrotron emission as the cause of a given source. The recent advances in radio imaging-spectroscopy can provide the spatial information required for investigating the emission mechanism while also supplying the spectral diagnostics that can be exploited when gyrosynchrotron emission is favored. In any case, the rare observations of type IVm associated with CMEs are a powerful tool to investigate the densities and magnetic field strength in these energetic events.

In recent years, new instruments have come online that, while not solar-dedicated, nevertheless can be used for occasional imaging spectroscopy of the Sun at high temporal, spectral, and spatial resolution. Two in particular have helped to renew interest in metric and decimetric studies of the radio Sun, the Low Frequency Array \citep[LOFAR;][]{2013A&A...556A...2V} and the Murchison Widefield Array \citep[MWA;][]{2013PASA...30....7T}. LOFAR operates in two frequency ranges, 10--90~MHz and 110--250~MHz, while the Murchison Widefield Array (MWA) observes between 80--300~MHz.  Both provide imaging spectroscopy of the radio Sun at high temporal resolution on an occasional basis. Several studies have been published in recent years that take advantage of these novel instruments to investigate both active and quiet Sun \citep[][e.g.,]{mann2018, zucca2018, sharykin2018, reid2018, McCauley2019, Mondal2020, rahman2020, sharma2018}.

In this paper we report the first analysis of solar data from the Owens Valley Radio Observatory Long Wavelength Array (OVRO-LWA), located near Big Pine, CA. At the time of observation, the array consisted of 288 dual-polarization dipole antennas which are optimized to minimize side-lobes; 256 residing in a 200 m diameter core, and the remaining 32 extending to maximum baselines of $\approx\,$1.6 km \citep{2019ApJ...886..123A,2019AJ....158...84E}, allowing it to spatially resolve the Sun in the frequency range 27--85 MHz with high spectral resolution. The spatial resolution is about 8.5\arcmin\ at 80~MHz.  The total bandwidth of the array covers this frequency range with 22 subbands, each comprising 109 24-kHz-wide channels at an operational cadence of either 9 or 13 s, although a 1-s snapshot mode is also available. The array uses the 512-input LEDA correlator \citep{2015JAI.....450003K}, which allows imaging the whole visible hemisphere at all times. Upon completion of an upgrade now underway, it will consist of 352 elements spanning a maximum baseline of $\approx$2.6 km, with a correspondingly improved spatial resolution of 5\arcmin\ at 80~MHz. The upgrade includes a dedicated solar mode with full-spectrum imaging at 0.1-s cadence and measurement of the flux density dynamic spectrum at 1-ms time resolution.

These ``first-light'' solar data were recorded during commissioning in 2015 to explore the capabilities of the array for solar radio observations, both at quiet times and during energetic events. We present our investigation of a complex event consisting of multiple bursts and sources, which occurred in association with a CME and a GOES soft X-ray (SXR) class M2.1 flare. In $\S$\ref{sec:Observations} we present the observations obtained from OVRO-LWA and the supplementary observations taken from other instruments. In $\S$\ref{sec:Analysis} we investigate the classification of each of the major burst sources and analyze the various features to identify the emission mechanism. We find that one of the sources, a moving type IV (IVm) is due to gyrosynchrotron emission and use spectral fitting to estimate the evolving physical parameters corresponding to the core of the CME. In $\S$\ref{sec:Discussions} we discuss our results and conclude with a discussion of what solar research will be possible with the instrument upgrade now underway.

\section{Observations} \label{sec:Observations}

\begin{figure}[b]\centering
\includegraphics[width=.7\textwidth, clip]{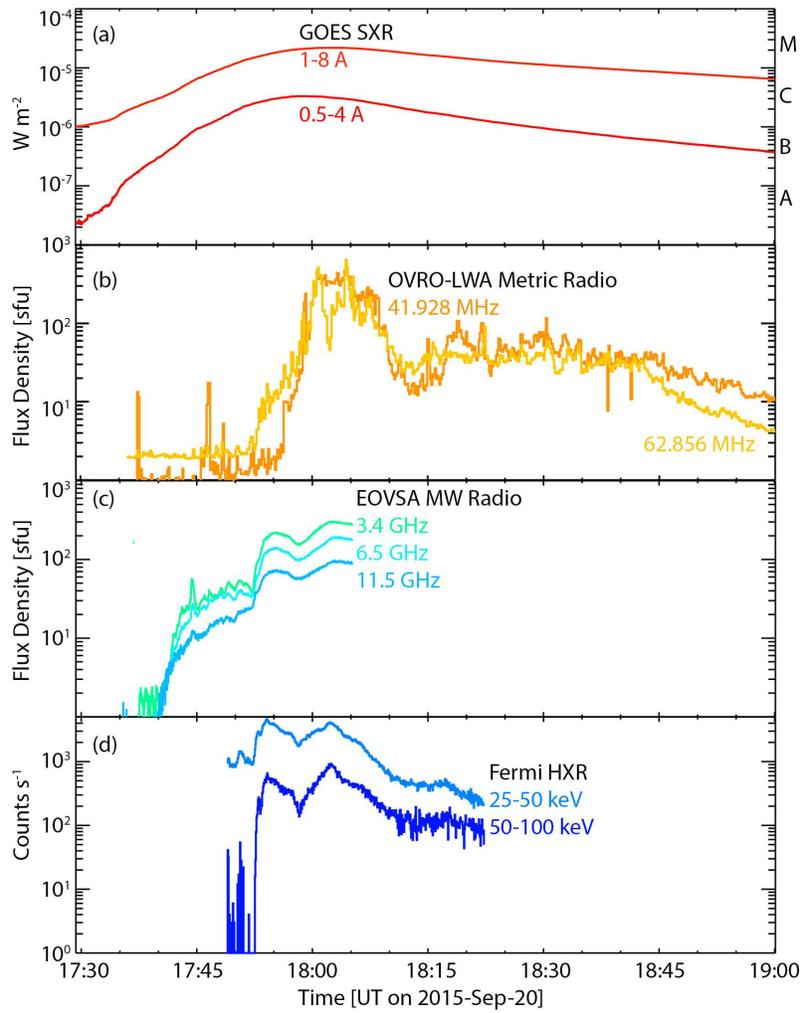}
\caption{\label{lightcurves} 
Radio and X-ray light curves showing the temporal development of the M2.1 solar flare on 2015 September 20. (a) GOES SXR. (b) OVRO-LWA metric flux density at two representative frequencies. (c) EOVSA flux density at three representative frequencies. (d) Fermi/GBM HXR count rate accumulated over two nonthermal energy ranges.}
\end{figure}

A GOES SXR class M2.1 flare and associated CME were observed on 2015 September 20 (SOL2015-09-20) in NOAA active region 12415, located at heliographic coordinates S19W52, near the south-west limb. The flare onset was at 17:32 UT (Figure~\ref{lightcurves}) and the SXR peak was reached at 18:03 UT, accompanied by a rather complex radio event recorded by OVRO-LWA also reaching peak emission around the same time as the SXR flux.  The SXR flux returned to GOES-class C1 (background) level several hours later, around 21:00 UT. The first signature of the white light CME (WL-CME) was detected at 18:12 UT in the field of view of the C2 coronagraph of the Large Angle and Spectrometric COronagraph \citep[LASCO;][]{Brueckner95} instrument onboard the Solar and Heliospheric Observatory \citep[SOHO;][]{domingo95}.

The event recorded by OVRO-LWA coincided with the commissioning observations for expansion of the array to the current maximum baseline of $\approx\,$1.6 km. This early stage of operations had three major effects on the data collected: 1) the observed bandwidth was reduced to a little less than 40\% of the total available; 2)the baseline frequencies were offset by 150 kHz, and 3) adequate understanding of the circular polarization calibration was not yet available, hence we do not attempt to use circular polarization information for the event. The first effect reduced the bandwidth to two available windows between 41--55 MHz and 61--69 MHz as opposed to the full bandwidth capability of 27--85 MHz and the second resulted in an offset in the position of all sources present in the sky, which we corrected. The third problem limits us to consideration of total intensity only. None of the above-mentioned drawbacks had any impact on the spectral and temporal resolution of the instrument during the event.

Observations from the LASCO C2 instrument covering 2.2--6 $R_\odot$ (distance from the solar disk center) were used to get context of the radio source with respect to the WL-CME, and also to provide electron density diagnostics. We also use data from the WAVES \citep{Bougeret95} radio spectrograph onboard the WIND spacecraft, where spectra from both RAD1 and RAD2 receivers was obtained. The frequency ranges between 20--1040 kHz and 1.075--13.825 MHz are covered by RAD1 and RAD2 respectively; the Expanded Owens Valley Solar Array \cite[EOVSA;][]{Gary2014} for microwave radio spectra that covers 1--18 GHz; and the US Air Force Radio Solar Telescope Network (RSTN) spanning between 25--180 MHz for meter-wave spectra to analyze the event and provide context among multiple wavelengths.

\subsection{Data Collection and Processing for OVRO-LWA} \label{subsec:Data Collection and Processing}
The flare-CME event was analyzed over a period of 120 min with images of 9-s cadence. Images during this period were inspected to identify distinct sources, understand their spatial configuration and describe their temporal behavior. Additional spectral analysis was done for certain times when the complexity of the burst was at minimum, to provide information on the emission mechanism of the burst. Since OVRO-LWA observes the whole sky at all times, to obtain the images for our analysis of the solar radio bursts, we first convert the data from the native output of the correlator to the Measurement Set (MS) format \citep{mcmullin07} using an in-house tool called ``dada2ms''. The data are further processed in the following 4 steps: 
\begin{enumerate}
\item \textit{Flagging and calibration}: Once the Measurement Sets are corrected for the aforementioned frequency offset, adopting the strategy outlined in \cite{anderson2018}, all identified bad antennas, channels and baselines are flagged and the data are calibrated using bright sources in the sky, e.g. Cyg A and Cas A. To minimize the presence of side-lobes, the bright sources are later removed from the all-sky maps (using a process known as ``peeling'') \citep{2008ISTSP...2..707M, 2018AJ....156...32E}.
\item \textit{Shifting the phase center} of the sky maps to the nominal position of the center of the Sun from the originally Zenith-centered calibrated maps.
\item  \textit{Cleaning} the sky maps and creating multi-frequency synthesis (MFS) FITS images using WSClean \citep{Offringa2014} for each subband. The available bandwidth during the time of the event consists of 9 subbands, each having 109 channels, 24~kHz wide, which are merged together to create 9 band-averaged MFS images at each time. All frequencies mentioned hereafter are the central frequencies of these subbands.
\item \textit{Flux calibration and correction for position offsets} caused by the ionosphere and instrumental effects: The radio galaxy Virgo A happened to be within 15 degrees of the Sun.  It is a strong source with a well-established flux density model for low frequencies \citep{Vinyaikin2016} and a precisely known position. We use this source to calibrate solar flux density and position. 
\end{enumerate}

\begin{figure}[t]\centering
\includegraphics[width=.5\textwidth, clip]{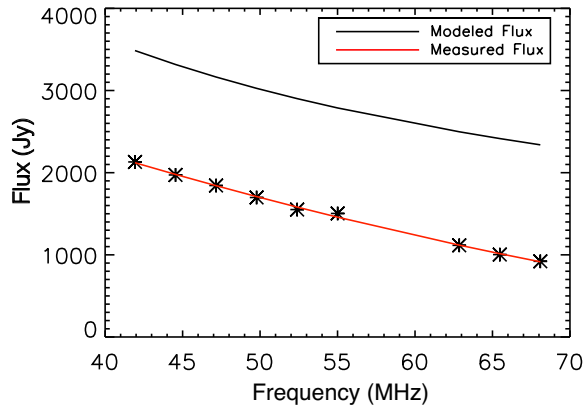}
\caption{\label{fluxfactor} 
Measured and modeled flux density of radio galaxy Virgo A. The well-established flux density model for low frequencies is taken from \cite{Vinyaikin2016} (solid line). The flux for Virgo A (asterisks) was measured at our 9-s cadence, averaged over a few minutes and fit to a quadratic polynomial shown by the red line.
}
\end{figure}

Figure \ref{fluxfactor} shows the measured, uncorrected flux density for Virgo A averaged over a few minutes, represented by asterisks fitted to a second-degree polynomial, along with the model flux density spectrum a factor of $\approx$2 higher. We use the model flux density and the polynomial fit to the measured flux of Virgo A, to obtain a flux factor as a function of frequency, which is then used to ``bootstrap'' the measured solar flux density to the corrected values. Flux correction for residual primary beam effects is also performed in this step.

Large scale fluctuations in the ionosphere can cause a refractive offset in position, which needs to be further corrected in order to measure the source centroid position with accuracy. Ionospheric refraction is greater at lower frequencies and falls quadratically with increasing frequency. It can be measured and corrected by referring to a nearby point source with a known position. As noted earlier, Virgo A is $\approx$15$^\circ$ from the Sun and so, is well-placed for use in correcting for refraction offsets.
The precisely known position of this source allows us to determine the offset in its observed position with respect to its true position as a function of frequency. We find a fairly small shift of 10$''$--30$''$ in right ascension (RA) increasing from the lowest to highest frequencies, and a larger shift of 320$''$--160$''$ in declination (DEC), which we correct. Temporal variations in the position of Virgo A were found to be within $\pm10''$ and have been ignored.  The measured offsets are indeed found to follow the expected quadratic pattern, which gives us confidence that the refraction due to the ionosphere has been successfully corrected.

\begin{figure}[t]\centering
\includegraphics[width=1.0\textwidth, clip]{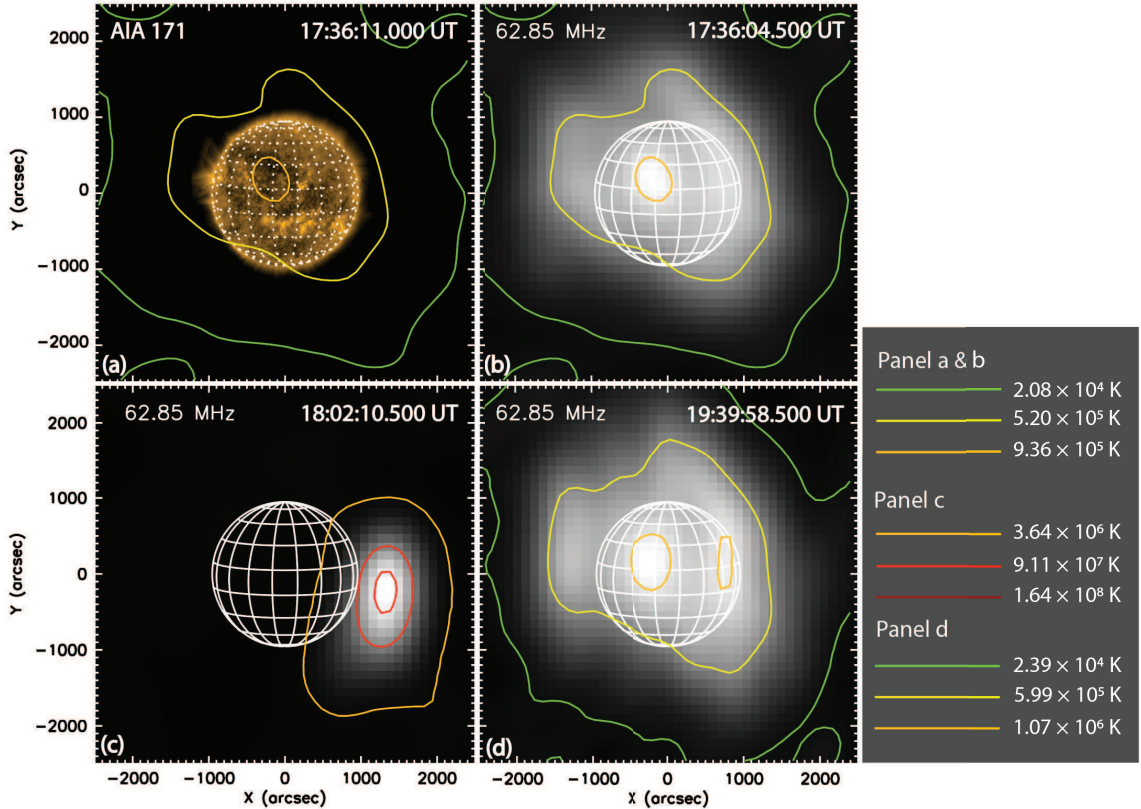}
\caption{\label{contextimage} Processed images of the Sun at different times during the event to show the dynamic range of the instrument. (a) AIA 171\AA\ image overlaid with quiet-Sun contours from panel b for comparison. (b) A representative quiet Sun image (gray scale) from a time before the event, with contour levels indicated in terms of $T_b$ in the box on the right. These represent 2$\%$, 50$\%$ and 90$\%$ of the maximum brightness temperature. (c) Same as b, for a time near the peak of the burst.  Now the 2\%, 50\% and 90\% contours represent much higher $T_b$ values. (d) The quiet Sun after the event, similar to the pre-event image in panel b, although the west limb $T_b$ is slightly elevated.
}
\end{figure}

Representative images of the Sun are shown in Figure \ref{contextimage} at 62.85 MHz for three different times during the event after the calibration, synthesis imaging, and offset correction process is complete. The solar grid is superimposed for scale along with a reference image from the Atmospheric Imaging Assembly (AIA) 171 \AA\ channel onboard the Solar Dynamic Observatory \citep[SDO;][]{lemen2012, boerner2012}. Each LWA image includes three contours representing 2$\%$, 50$\%$ and 90$\%$ of the total brightness temperature $T_b$ demonstrating the wide dynamic range of the instrument.

\subsection{Event Overview} 

\begin{figure}\centering
\includegraphics[width=.65\textwidth, clip]{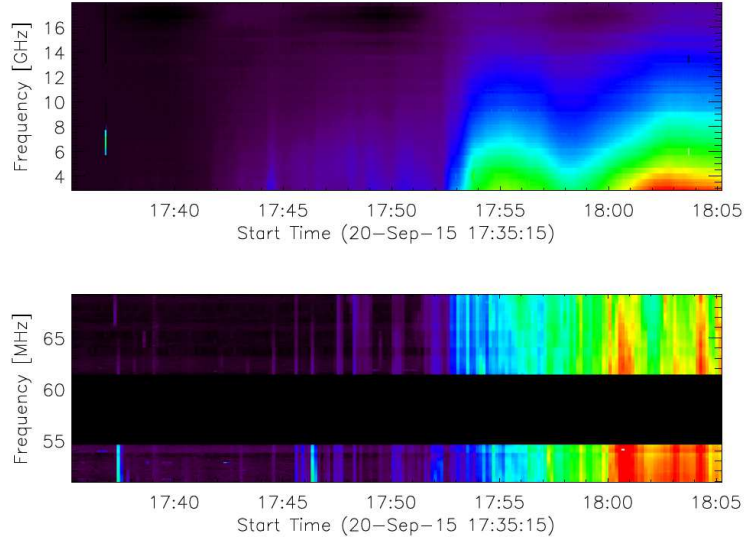}
\caption{\label{EOVSAds} 
LWA (bottom) dynamic spectrum in context with the corresponding dynamic spectrum from EOVSA (top) obtained at some 2-orders-of-magnitude higher frequency in microwaves.}
\end{figure}

As shown in Figure~\ref{lightcurves}, the time profiles at a wide range of wavelengths and energies are similar. The microwave flux density from EOVSA, which is due to gyrosynchrotron emission, was strongest at lower microwave frequencies (3.4 GHz) and shows nearly the same but weaker time profiles at higher frequencies.  The dynamic spectrum from EOVSA is shown in Figure~\ref{EOVSAds} up to 18:05~UT, when observations ended, compared with the OVRO-LWA dynamic spectrum over the same time period.  
The two episodes of enhanced activity in microwaves, peaking at 17:55 UT and again at 18:03 UT, are also seen in HXR emission by the Gamma-ray Burst Monitor aboard Fermi (Fermi/GBM; Figure~\ref{lightcurves}d), but the microwave times are delayed 1--2 minutes relative to the peak times of nonthermal HXR emission, suggesting relatively strong trapping of the microwave-emitting electrons. Note that gyrosynchrotron microwave spectra typically peak between 5--10~GHz \citep{1975SoPh...44..155G} while the peak microwave frequency for this event seems to extend below the lowest observed frequency in EOVSA (2.9~GHz) at the time.  This suggests a relatively low magnetic field strength in the source, and hence a greater height for the emission, than is typical for microwave bursts.

\begin{figure}\centering
\includegraphics[width=.95\textwidth, clip]{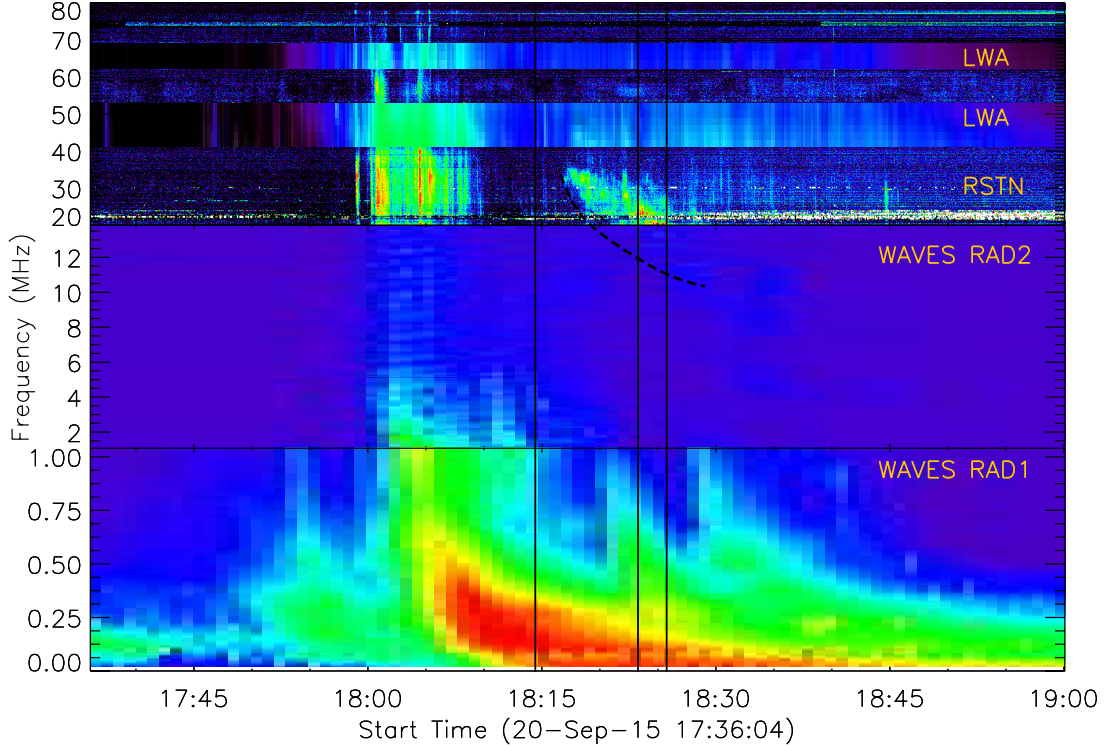}
\caption{\label{dynspec} 
Composite dynamic spectrum from WIND/WAVES RAD1, RAD2, RSTN with an overlay from OVRO-LWA.
 }
\end{figure}

We extract dynamic spectra from the WIND/WAVES and RSTN radio instruments, which cover frequencies both above and below the observing range of OVRO-LWA, for contextual understanding of the observed radio bursts. RAD1 and RAD2 receiver bands from WAVES onboard the WIND spacecraft cover the frequency range between 20--1040 kHz and 1.075--13.825 MHz with 256 channels each and a bandwidth of 3 kHz and 20 kHz respectively. The RSTN spectrograph covers 25--180 MHz, overlapping with the OVRO-LWA range. Figure \ref{dynspec} shows a composite dynamic spectrum with the lower frequencies from WIND/WAVES RAD1 and RAD2, and the higher frequencies from RSTN.  The available bands in OVRO-LWA, with their far higher signal to noise ratio, are inserted to replace the RSTN spectrum in those bands. The three vertical black lines mark times to be discussed in section~\ref{Emission mechanism}. The peak of radio emission occurs at a similar time as the soft X-ray peak at 18:03 UT and is associated with a dense group of type IIIs.  Continuous emission is seen in the decay phase consistent with a type IV burst. There is also a signature of a type II burst at lower frequencies, seen in RSTN and extending into RAD2 frequencies (whose leading edge is marked by dashed-line in Figure \ref{dynspec}).  We searched for an OVRO-LWA counterpart by extrapolating the type II emission to higher frequencies and earlier times, but did not find any clear signature despite OVRO-LWA's high sensitivity.  This suggests that the type-II-burst-emitting shock had not formed until later and at greater heights.

LASCO-C2 images with a 12-min cadence were obtained between 18:12--19:12 UT and processed to enhance the CME features using monthly background subtraction, median filtering, occulter masking and Normalized Radial Graded Filtering \citep[see][for details]{Morgan2006}. The upper row of panels in Figure~\ref{overview} show LASCO-C2 images at four times during the event.  In the first image at 18:00 UT, the CME had not yet reached above the C2 occulter at 2.2 $R_\odot$. Its first appearance was at 18:12 UT (see contours in the left column of Figure~\ref{movingsource}). A height-time plot of the CME leading edge and core is shown in Figure~\ref{htplot}, which indicates a relatively constant velocity (1240 km s$^{-1}$) that extrapolates back to 1 $R_\odot$ around 17:57 UT.  This suggests an association with the first of microwave and HXR peaks around 17:55 UT.  No data were available for this observation window from COR1 and COR2 coronagraphs onboard the STEREO satellites. The closest in time STEREO-A COR2 image was at 19:26 UT, and because the Earth--spacecraft angle was 170.6 degrees, its view was almost directly behind and provides little additional information about the plane of the CME.

\section{Burst Identification \& Analysis} \label{sec:Analysis}
\subsection{Burst Identification} \label{Burst Identification}

\begin{figure}[t]\centering
 \includegraphics[width=1.\textwidth, clip]{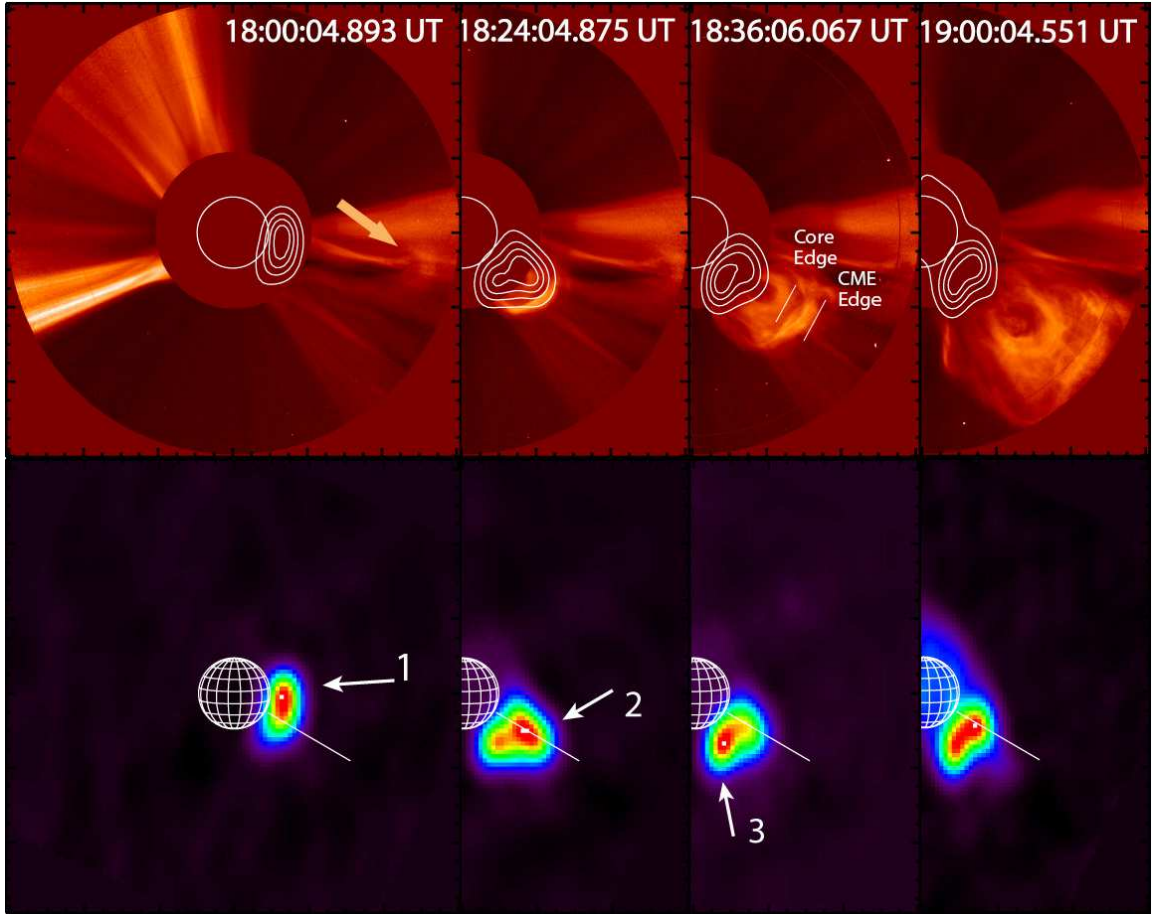}
 \caption{\label{overview} 
 \textit{Top row:} Evolution of the flare-CME event in WL recorded by LASCO-C2 at four selected times constrained by LASCO's 12-min cadence. The yellow arrow in the left panel points to a pre-existing CME discussed in the text. OVRO-LWA 20\%, 40\%, 60\% and 80\% brightness temperature contours are overlaid for reference.  The third panel shows example locations of the leading edge of the CME and the core used for height-time analysis. \textit{Bottom row:} OVRO-LWA maps at 62.86~MHz, at the corresponding times for comparison.  White arrows mark three source-regions of interest (SRs) described in the text. The white lines mark the slit along which height-time measurements are made for Fig.~\ref{htplot}.
 A movie of the complete event recorded between 17:36 UT--19:36 UT is available. The left side panels show the LASCO-C2 images with OVRO--LWA contours overlaid. The right side are the OVRO-LWA maps recorded at 41.93~MHz in the top panel and at 62.86~MHz in the bottom panel. The LASCO-C2 images have a cadence of 12 min while the OVRO-LWA images have a cadence of 9 s. The real time duration of the animation is 111 seconds.}
 \end{figure}

OVRO-LWA images of the Sun are created at each time in the 9 available frequency-subbands to create a movie between 17:36-19:36 UT, beginning with the onset of the flare and ending with the decay phase (See movie in Figure \ref{overview}). Upon carefully examining the complex event as seen by OVRO-LWA and LASCO-C2 (C2 hereafter) WL data, we identify distinct times and source-regions of interest (TOIs and SRs hereafter) based upon important changes in burst position and/or apparent motion. We use the term source-region (SR) and source interchangeably hereafter. Figure \ref{overview} gives an overview of the event with WL images as recorded by C2 in the top panels, and corresponding OVRO-LWA maps in the bottom panels. OVRO-LWA brightness temperature contours at 62.86~MHz (one of the higher frequencies available) are overlaid on the C2 maps. SRs 1, 2 and 3 are identified with white arrows at those times when each becomes dominant at 62.86 MHz. The identified SRs represent three different burst sources spatially distributed within the erupting region with possibly different underlying mechanisms. The order of appearance and dominance of burst emission in the SRs varies with frequency and time in a complicated manner due to their different spectral and temporal behavior.  However, source 2, which is well aligned with the axis of the CME, shows an outward motion at least during some of the time, as described further below. 

A concise description of the dominant sources during the TOIs is noted in Table \ref{table1}. There are signatures of numerous distinct bursts in SR 1, lasting less than our 9-s cadence, observed first in lower frequencies and later over the whole band, immediately after the onset of the flare at 17:32 UT and continuing for about 15 minutes (associated with a slowly increasing, but weak level of activity in microwaves visible in Figure~\ref{EOVSAds}). Fairly continuous broadband emission, but fluctuating in brightness, comes mainly from SR 2 until a few minutes before the peak of the emission at $\approx$18:03 UT. Relatively stationary pulsations from SR 1 become dominant at lower frequencies during this peak time, while at higher frequencies emission from both SR 1 and 2 are comparable. Around 18:18 UT, SR 3 first appears. From 18:18--18:23 UT sources 1 and 3 fluctuate but SR 2 exhibits steady emission smoothly varying both in frequency and time and a distinct outward movement across the whole band, reaching as high as 3 $R_\odot$. 
\begin{deluxetable*}{ccCrlc}
\tablecaption{\label{table1}Overview of the event with respect to the identified source regions (SRs) at different times, and the respective features observed}
\tablecolumns{4}
\tablenum{1}
\tablewidth{0pt}
\tablehead{
\colhead{Time of interest} &
\colhead{Dominating SR} &
\colhead{Dominating SR} & \colhead{Dominant Features Observed}  \\
\colhead{(UT)} & \colhead{(Low $\nu < 55~{\rm MHz}$)} &
\colhead{(High $\nu$)} 
}
\startdata
17:39--17:47 & 1 & 1  & Sporadic short-duration bursts$^a$ \\
17:49--17:54 & 2 & 2 & Continuous Emission \\
18:00--18:09 & 1 & 1,2 & Stationary pulsations \\
18:18--18:23 & 1,2,3 & 2,3 & Outward movement in 2 \\
 \enddata
\tablenotetext{a}{Mainly type III bursts}
\end{deluxetable*}
\subsection{Burst Analysis} \label{Burst Analysis}

\textit{SR 1}: The examination of the movie alongside the dynamic spectrum suggests that emission in SR 1 is likely associated with type III radio bursts. Several distinct type IIIs are observed in the dynamic spectrum between the onset and the peak time of the flare, and a dense group of type III bursts identified from the dynamic spectrum appeared from both SR 1 and 2 during the peak. The SR 1 burst location spatially aligns with the position angle of a previous, slow and narrow CME, first observed with C2 at 15:48 UT. The remnant structure of this previous CME is shown by the yellow arrow in Figure~\ref{overview}. The material from this previous CME has moved out to $\approx\,$5.38 \(R_\odot\) by the time we observe type III activity in the region.  One possible explanation for the existence of type III bursts at this location is that there may be some interaction between this pre-existing magnetic structure remnant from the earlier CME and the initial stages of the M2.1 flare. Alternatively, \cite{2000ApJ...530...1049} have shown that a class of hectometric type III bursts observed over an extended duration and temporally associated with decimetric and metric radio emission may be produced by electron beams ejected primarily from the flare site itself. Such bursts are classified as complex type III bursts due to their complex intensity profiles.

\begin{figure}[hb]\centering
\includegraphics[width=.8\textwidth, clip]{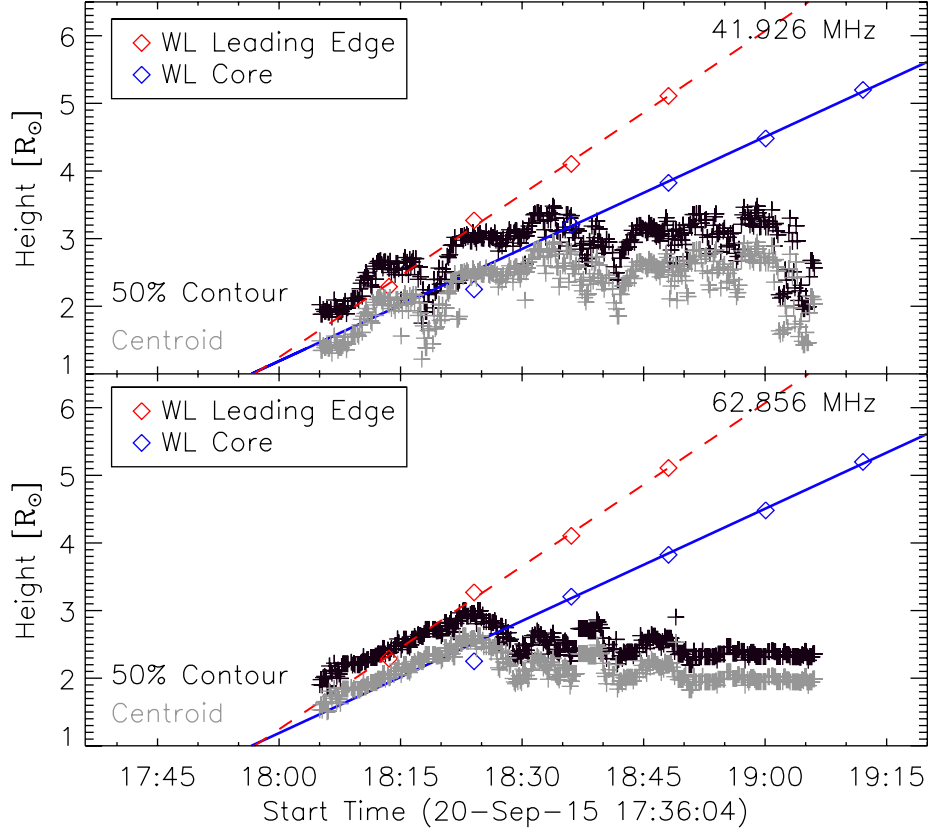}
\caption{\label{htplot} 
Composite height-time plot of the radio source SR 2 at two frequencies, 41.92~MHz (top) and 62.86~MHz (bottom) with WL CME leading edge (red) and Core (blue), overlaid. The black symbols mark the position of 50\% contour of the radio source, and gray points show the source centroid estimated by subtracting the beam half-width.}
\end{figure}

\textit{SR 2}: \textit{Height-Time Analysis}: As mentioned above, it is in SR 2 that we observe a steady outward movement of the source. This motion is investigated in detail by performing a manual height-time analysis of the radio source at all the available frequencies to estimate the velocity of the source and then compare it with the speeds of the CME leading edge and the core. For the purpose of height-time analysis of the radio source, we chose a direction aligned with the observed outward motion (shown by the white lines in the bottom row of Figure \ref{overview}), which is only slightly non-aligned with the direction of CME motion. To measure the outward motion of the radio source, we do not use the peak or centroid locations, both of which were affected by the intermittent brightenings of overlapping source SR 1, but rather we use the position of the apparently more stable 50\% contours of SR 2 (the intersection of this contour with the axis line) at each time as a measure of this outward motion. The 50\% contour height is shown by the black symbols in Figure \ref{htplot} at two different frequencies.  To obtain a rough proxy for the source centroid position, which cannot be directly measured due to the above-mentioned confusion with SR 1 \& 3, we subtract the synthesized beam (point-spread-function) half-width at each frequency, under the assumption that the source is unresolved. In cases where the source is resolved, this proxy will slightly overestimate the height of the centroid, but it is sufficient for our purposes here.  This proxy for the centroid position is plotted in Figure \ref{htplot} as gray symbols. Heights of the outward moving leading edge and core of the WL CME are also measured in a similar manner (one example is shown in the third panel of the upper row of Figure~\ref{overview}). The positions are recorded where the outer edge of the core (blue diamonds in Figure \ref{htplot}) and the leading edge (red diamonds) intersect the axis at each time. The same points are plotted in both upper and lower panels of Figure~\ref{htplot}.  The blue core points are fit with a constant velocity (775 km s$^{-1}$). The first point (at 18:24 UT) was excluded while performing this fit, since the height of the core at that time is uncertain due to its proximity to the occulter. The fit is extrapolated backward to highlight that the CME core velocity matches the gray radio centroid points quite well.

A smooth outward motion is evident at higher OVRO-LWA frequencies in Figure \ref{htplot}, while the lower frequencies show rather erratic changes, due to greater source confusion, although also having an outward-moving trend. Although an apparent motion is clear from the height-time plots, visible outward motion in the movie is most apparent only for a short of period of time between 18:18--18:23 UT when only SR 2 was visible and undisturbed by the other sources. Based on the outward motion of the source, along with its clear match to the WL CME core position and speed, we classify the SR 2 as a type IVm (moving type IV) source.  
\\~\\

\textit{SR 3}: SR 3 aligns well with the southern flank position of the CME, and the source becomes dominant during the decay phase of the radio event, while the source centroid over all observed frequencies appears to be spatially coincident. Plausible mechanisms may include stationary type IV plasma emission from particles accelerated in this southern flank region. However, due to the lack of stereoscopic observations and polarization information from the OVRO-LWA, any additional conjecture regarding the morphology and underlying mechanism is not justified. 

\subsection{Moving Type IV Emission Mechanism} \label{Emission mechanism}
 \begin{figure}[!hb]\centering
 \includegraphics[width=1.\textwidth, clip]{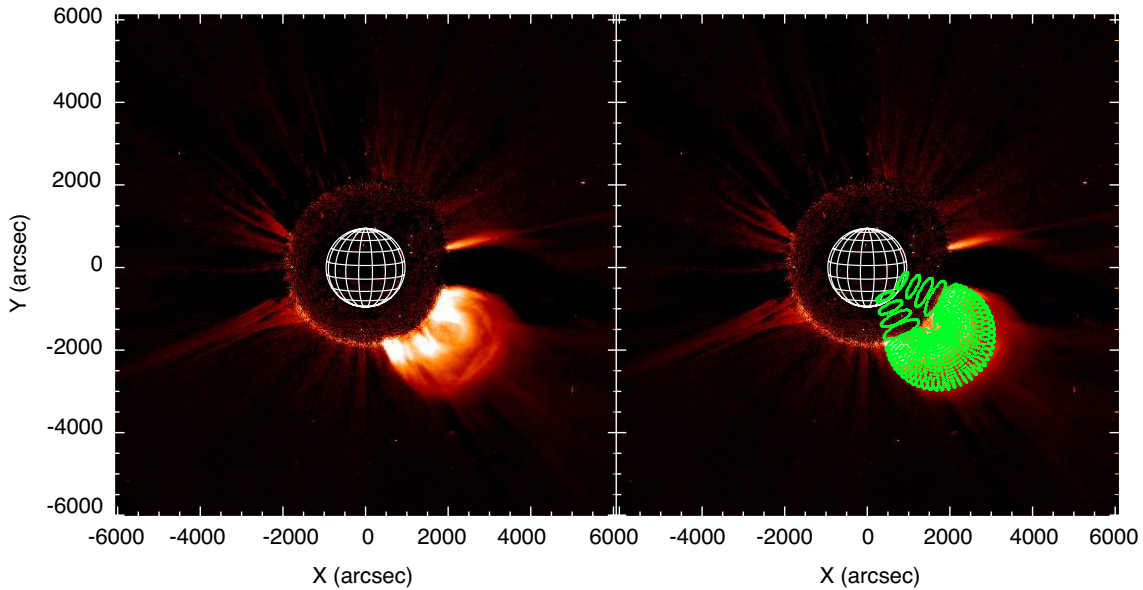}
 \caption{\label{GCS} 
 \textit{Left}: Reference image of the C2 WL CME at 18:36 UT used to create a CME model by the GCS reconstruction method. \textit{Right}: Wire-frame CME in green.
 }
 \end{figure}
\begin{figure}[b]\centering
\includegraphics[width=.95\textwidth, clip]{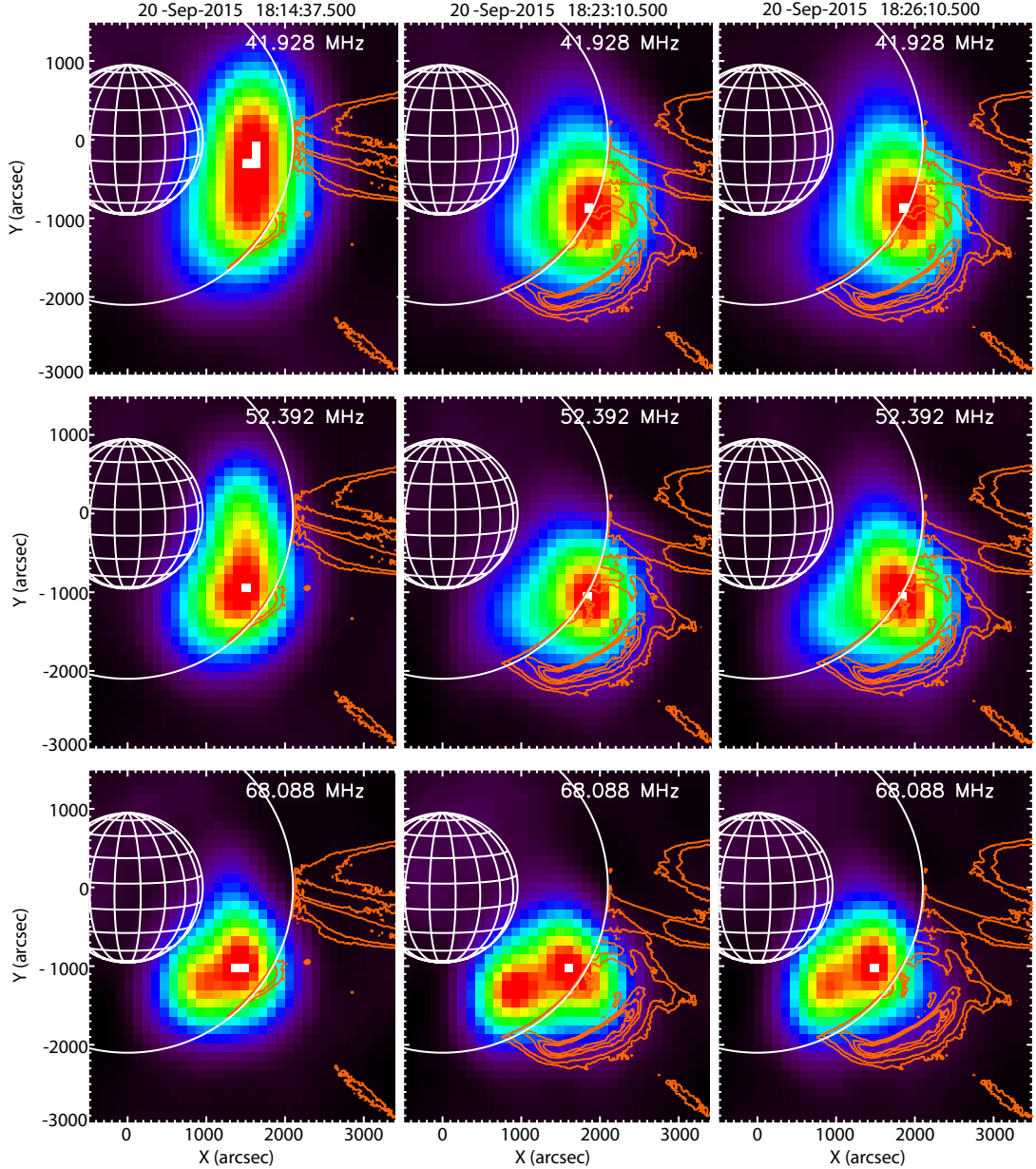}
\caption{\label{movingsource} 
Overview of the outward moving radio source in SR 2 as seen by OVRO-LWA. WL CME contours from C2 are overlaid in orange to show the relative movement of the source with respect to the CME. Time is increasing as we move from left to right and frequency increases as we move from top to bottom. The C2 contours in column 2 \& 3 are the same (taken at 18:24 UT), the only frame closest in time with respect to the OVRO-LWA images. C2 contours in column 1 are taken at 18:12 UT. The white solid circle marks the C2 occulter at 2.2~$R_\odot$.}
\end{figure}

\begin{figure}\centering
\includegraphics[width=.95\textwidth, clip]{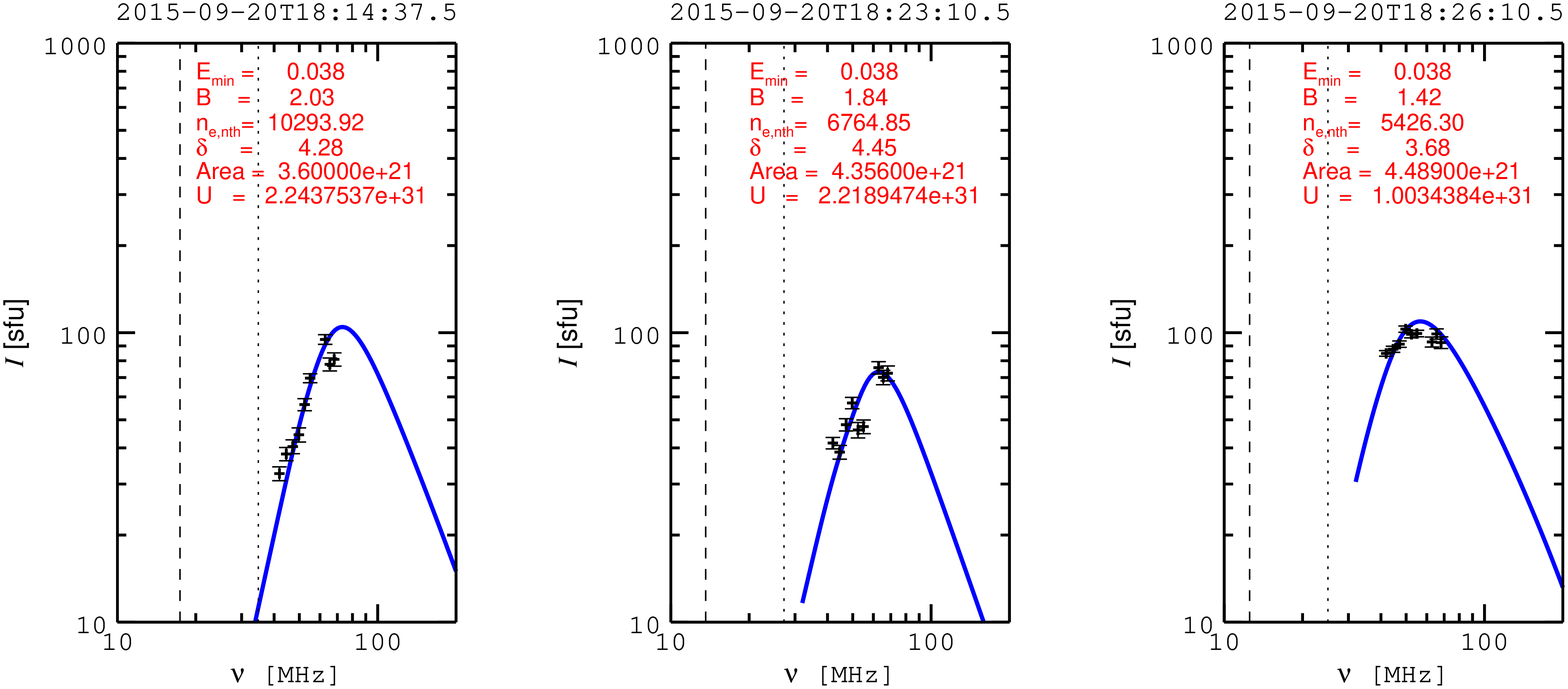}
\caption{\label{lwaspectra} 
Spectral fits for the moving source in SR 2 at three times during the observed outward motion. From the ambient density used in the fits, the dashed and dotted vertical lines show the corresponding plasma frequency, $\nu_p$ and its second harmonic, 2$\nu_p$ respectively at each time, relative to the observed emission.}
\end{figure}

A number of studies \citep{Bastian2001, Maia07, Tun13, bain2014} have shown that continuum emission associated with CMEs can be due to gyrosynchrotron radiation.  While \cite{Maia07} \& \cite{Bastian2001} were able to rule out plasma emission owing to the well defined radio--CME loop morphology that was spatially coincident over multiple frequencies, we lack that advantage since our source is not spatially resolved. Therefore, we must rely on other arguments similar to those by \cite{Tun13} and \cite{bain2014}. The fact that the emission we observe varies smoothly over frequency and time for the duration of the visible outward motion, together with the rather large height of SR 2 ($> 2.5 R_\odot$ at 65~MHz), suggests gyrosynchrotron emission for our event as well. To investigate further whether plasma emission can be ruled out, we examine the density using constraints from the LASCO-C2 observations. It is noted that the limited resolution along with the source confusion restricts us to only average over the inhomogeneous structure. Any physical parameters derived further are only used as representative values and apply to the source only in an averaged--sense. 

We start by performing a graduated cylindrical shell (GCS) reconstruction of the CME \citep{thernisien06}. The model assumes a self-similar expansion of the CME, integrated with a flux-rope morphology. We note here that due to the lack of simultaneous stereoscopic observations we get only one viewpoint from LASCO, so we lack a firm measure of the plane-of-the-sky angle of the CME, on which the density depends.  However, fitting a single point of view can still offer a useful constraint on the density. The free parameters in the model such as CME height, tilt-angle of the source-region neutral line, angular width and aspect ratio are used to create a wire-frame model for the CME, shown in Figure \ref{GCS}. For plane-of-the-sky angle we take the initial values of the heliographic coordinates for the source region from SDO/AIA measurements, which are then adjusted to match the observations. Such adjustments are sometimes necessary given the fact that CME deflections are commonly observed due to several influencing factors like background coronal magnetic topology, streamers, and fast solar wind from coronal holes \citep[see][]{macqueen86, cremades05}.  The model is then used to generate a synthetic brightness image of the CME using a ray-tracing renderer, based on Thompson scattering equations. This process is performed for C2 images recorded at 18:24 UT \& 18:36 UT, and the density estimates for the leading edge of the CME are then adjusted to give synthetic brightness that matches with the observed brightness. Although the leading edge is a convenient place to use for scaling the GCS model, the scaled model is interpreted as applying to the entire CME including its core (represented by the inner fold of the croissant-shape) where the moving radio source is located.
We exploit the fact that the core and the leading edge of the CME are comparable in brightness (within a factor of 2) to get a proxy for density in the core. 
Using this process we obtain the value of density for the leading edge of the CME at 18:24 UT to be \(N_e\) $\approx$ \( 2.2\times10^6\,{\rm cm}^{-3} \) corresponding to a plasma frequency \textup{$\nu_p$} of 13.32 MHz. The density obtained at 18:36 UT is $\approx$ \( 6.2\times10^5\,{\rm cm}^{-3} \) corresponding to \textup{$\nu_p$} = 7.07 MHz. We use these estimates when discussing our spectral fitting results below.

\textit{Spectral Fitting}: Fast gyrosynchrotron codes \citep{fleishman10} are used to calculate the gyrosynchrotron emission based on a homogeneous source, which serves as the basis for performing spectral fitting at three different times (shown by vertical black lines in Figure \ref{dynspec}) during the observed outward movement of the source, to estimate its evolving physical conditions. We chose times when there is a clear dominance of the source emission in SR 2 such that it can be clearly separated from bursts in SR 1 \& 3 at most of the available frequencies. Figure \ref{movingsource} shows an overview of the outward moving source at three different frequencies and the chosen times, with the WL CME contours from C2 overlaid (orange) for reference.  The white solid line marks the C2 occulter. Note that the C2 contours in columns 2 \& 3 in the figure are the same (taken at 18:24 UT), since that is the frame closest in time to the OVRO-LWA images. C2 contours in column 1 are taken at 18:12 UT.  The 50\% contours of the source are then used as a measure of the convolved source size. We further deconvolve the source using the synthesized beam, to obtain an estimate of the source area \citep{wild1970}.  As mentioned above, we average over an inhomogeneous source while estimating the physical conditions in the region. Consequently, we also assume an isotropic electron distribution for the spectral fits. An extrapolation of the density estimates made using the GCS reconstruction and the source area are fed into the codes and the free parameters viz., \(E_{min}\), the minimum energy of the electrons in MeV, $B$, the magnetic field strength in G, \(n_{e,nth}\), the non-thermal electron density in cm$^{-3}$, and $\delta$ the power-law index are adjusted to match the observed spectra. Spectral fits for these times are given in Figure \ref{lwaspectra}, with text in each panel giving the fit parameters, \(E_{min}\), $B$, \(n_{e,nth}\), $\delta$,source area and $U$, where $U$ is the magnetic energy density in erg cm$^{-3}$. The dashed and dotted vertical lines show the plasma frequency, $\nu_p$ and its second harmonic, 2$\nu_p$ respectively at each time. These are slightly higher than, but close to the values we determined from the GCS fitting of the CME brightness.  The spectral peaks are well above these plasma-frequency limits, further supporting the interpretation that the emission is due to the gyrosynchrotron mechanism.  The results of the fits suggest that both accelerated electron density and magnetic field strength decline as the source expands outward, while the power-law index of the electrons hardens.
\section{Discussions} \label{sec:Discussions}

We present first-light observations recorded by the OVRO-LWA of a flare and CME associated with a rather complex event at metric wavelengths. The observations were made during the first 24 h of commissioning observations for the expansion of the array, which compromised the frequency coverage and polarization capability, but nevertheless provided sufficient imaging spectroscopy to allow new insights into the rarely observed moving type IV (type IVm) phenomenon. A detailed analysis of the event is performed to characterize the multiple burst types observed and isolate the times when the emission was dominated by the moving source. We examined the relationship of the radio emission to a WL CME observed with the LASCO-C2 instrument, with context radio data from RAD1 and RAD2 receivers on board the WIND spacecraft, RSTN and EOVSA. We identify 3 different source-regions of interest in the complex radio event.
\begin{enumerate}
\item SR 1 appears to align well with a region associated with a previous CME, observed at 15:48 UT. The type IIIs observed in this region during the peak phase of the flare may have resulted from turbulent interaction between the flare and some pre-existing magnetic structure from the earlier CME. 
\item The broadband continuum emission in SR2 is classified as a moving type IVm burst. There exists a clear agreement between the outward movement of the source centroid in SR 2 and the corresponding movement of the WL CME core that can be seen on comparing the OVRO-LWA and LASCO height-time plots. The velocity of the CME leading edge from CME catalogues (maintained by the Coordinated Data Analysis Workshops (CDAW) Data Center) is given to be 1240 km/s. The position of the radio source from the height-time plots of OVRO-LWA suggests a lower velocity of ${\approx}$775 km s$^{-1}$, appropriate to the expected lower speed of the CME core. We further perform a GCS reconstruction of the CME to constrain its density in the volume. The corresponding plasma frequency is somewhat lower than the frequency of the observed emission, but given the uncertainties due to such issues as density inhomogeneity, angle of the CME to the plane of the sky, and line-of-sight depth, we cannot completely rule out the possibility of plasma emission as the underlying mechanism based on density alone. However, the smooth variation of SR2 in frequency and time strongly argue for gyrosynchrotron emission as the preferred mechanism.  Under this assumption, we fit gyrosynchrotron spectra to the observations to obtain estimates of the physical parameters in the burst as the source evolves. The EOVSA dynamic spectrum suggests a reservoir of microwave-emitting electrons at an unusually large height early in the event, which could potentially serve as a source of particles escaping into the CME core region.

There have only been a few studies that attempted to estimate the physical parameters of the plasma from type IV bursts using imaging-spectroscopy, most of them reporting observations at higher frequencies \citep{Bastian2001, Maia07, Tun13, bain2014, Mondal2020}. See the summary compiled by \cite{Mondal2020}, their table 3. The estimates for the magnetic field vary widely, with several reporting $B\approx10$~G at similar heights compared to our lower value of order 1--2~G.  Only the \cite{Bastian2001} measurements are this low. This variability may be real, and simply reflect different conditions in different events.  Additionally, \cite{Bastian2001} and \cite{Maia07} report these values in radio--CME loops while the event observed in the current study and the ones reported by \cite{Tun13} and \cite{bain2014} are associated with the core of the CME. Conversely, power-law index, $\delta$, is found in a narrow range of 3.5--5 in the previous studies, and our range of 3.7--4.5 is no exception.  The nonthermal density, \(n_{e,nth}\) is particularly variable in these studies, ranging from $2\times10^2$--$2\times10^6$ cm$^{-3}$, but this depends greatly on the value of $E_{\rm min}$ used for the estimate. Our values of 5000--10000 cm$^{-3}$ are well in line with these.

\item Although there is a signature of a type II burst at lower frequencies observed in RSTN and WAVES Rad2 during the period of the outward motion, no such feature is visible in the OVRO-LWA spectrum at the earlier time expected by extrapolation to higher frequencies. It is interesting that the frequency range of type II emission seen in the RSTN data, drifting from 35 to 20~MHz over this time, closely matches the 2nd harmonic plasma frequency corresponding to both our CME leading-edge density estimate (26.6~MHz at 18:24~UT) and the thermal density used in our gyrosynchrotron fits (dotted lines in Figure~\ref{lwaspectra}). If we had had the full frequency-coverage for this event that OVRO-LWA is capable of, we would have been able to image both the gyrosynchrotron and type II emission simultaneously. 

\item The continuous but time-variable broadband emission observed during the decay phase of the flare, associated largely with SR 3, shows a spatial alignment with the CME southern flank position. Plasma emission from particles associated with the flank of the CME is a plausible explanation. The CME flank is the location of shocks causing metric type II bursts in some models \citep[e.g.][]{2007A&A...461.1121C}, but again we find no clear type II spectral feature in the OVRO-LWA data. Recent studies have also shown evidence of ``herringbones'' observed at the CME flanks \citep{morosan19}, although the high cadence for this observation does not allow us to see any fine structures that may be present in the dynamic spectrum. Alternatively, this continuum source may be due to shock particles accelerated elsewhere and transported to SR 3. 
\end{enumerate}

A second round of expansion of the OVRO-LWA instrument is now underway that will provide a 50\% improvement in spatial resolution and include solar-dedicated observing modes specifically designed for solar science.  This includes a solar-dedicated beam synthesized from its 352 antennas to provide full spectral coverage of the total flux from the Sun at 1~ms time resolution, and a solar-dedicated pipeline-imaging mode that will correlate data from 48 of its 352 antennas at a 100~ms cadence for high-resolution imaging spectroscopy in Stokes I and V. Slated for completion by 2021, this upgraded OVRO-LWA instrument will provide a powerful new tool to study meter-wavelength emission of both bursts and the quiet Sun during the coming solar cycle.

\acknowledgments

S.C. thanks Robin Colaninno and Angelos Vourlidas for their helpful discussions in understanding the CME morphology. S.C. also thanks Camilia Scolini for sharing her insights on the GCS reconstruction method. This work was supported in part by the NSF grants AST--1615807, AGS--1654382, and AST--19010354 and by NASA grants 80NSSC18K1128 and 80NSSC17K0660 to New Jersey Institute of Technology. G.H. acknowledges support by NSF grants AST--1654815, AST--1828784. L.G. acknowledges support by NSF grants PHY--0835713, OIA--1125087, AST--1106059 and AST--1616709. The CME catalog is generated and maintained at the CDAW Data Center by NASA and The Catholic University of America in cooperation with the Naval Research Laboratory. SOHO is a project of international cooperation between ESA and NASA.

\end{document}